\documentclass[reqno]{article}
\usepackage{pifont}
\usepackage{amsthm}
\usepackage{color}
\usepackage{amssymb}
\usepackage{mathrsfs}
\usepackage{amsmath, amsfonts, vmargin, enumerate}
\usepackage{graphics}
\usepackage{verbatim}
\usepackage{amsthm}
\usepackage{latexsym,bm}
\usepackage{euscript}
\usepackage[noblocks]{authblk}
\usepackage{geometry}
 \makeatletter

\numberwithin{equation}{section}

\newcommand\USU{Department of Mathematics and Statistics, \\ Utah State University, Logan, UT 84322-3900, USA}
\newcommand\CH{School of Mathematics and Statistics and Hubei Key Laboratory of Mathematical Sciences,\\ Central  China Normal University, Luo-Yu Road 152, Wuhan, 430079, P.R. China}

\begin{document}

\title{On the (in)validity of the NLS-KdV system in the study of water waves
\footnotetext{{\it Corresponding author:} nghiem.nguyen@usu.edu.}}
\author{
CHUANGYE LIU \\
{\small\CH}\\ \vspace{3mm}NGHIEM V.NGUYEN \vspace{3mm}\\{\small \USU}
}

\date{}
\maketitle
\markboth{Chuangye Liu  and  Nghiem V.Nguyen}%
{On the (in)validity of the NLS-KdV system in the study of water waves}
\begin{abstract}
It is universally accepted that the cubic, nonlinear Schr\"odinger equation (NLS) models the dynamics of narrow-bandwidth wave packets consisting of short dispersive waves, while the Korteweg-de Vries equation (KdV) models the propagation of long waves in dispersive media.  A system that couples the two equations seems attractive to model the interaction of long and short waves and such a system has been studied over the last few decades.  However, questions about the validity of the system in the study of water waves were raised in our previous work where we presented our analysis using the fifth-order
KdV as the starting point.  In this paper, these questions are settled unequivocally as we show that the NLS-KdV system or even the linear Schr\"odinger-KdV system cannot be resulted from the full Euler equations formulated
in the study of water waves.
\end{abstract}
\section{Introduction}

For the last few decades, considerable attention ({\em e.g.} \cite{albert, pava1, pava2, chen, corcho, dias}) has been devoted to the study of the following system, which has been termed the cubic nonlinear Schr\"odinger-Korteweg-deVries (NLS-KdV) system:

\begin{equation}
\left\{
\begin{matrix}
\begin{split}
i u_t + u_{xx} + a|u|^2u &= -buv,\\
v_t + c vv_x + v_{xxx} &= -\frac{b}{2}(|u|^2)_x,
\end{split}
\end{matrix}
\right.
\label{NLS-KdV}
\end{equation}
where $x,t \in \mathbb R$, $v(x,t)$ is a real-valued function, $u(x,t)$ is complex-valued and $a,b,c$ are real constants. In the context of water waves, the NLS-KdV system was originally introduced by Kawahara \textit{et al.} \cite{KSK} in the form
\begin{equation}
\left\{
\begin{matrix}
\begin{split}
i\bigg(\frac{\partial u}{\partial t_2} + k \frac{\partial u}{\partial x_2}\bigg) + p \frac{\partial^2 u}{\partial x_1^2} &= q uv,\\
\frac{\partial v}{\partial t_3} + \frac{\partial v}{\partial x_3} + \frac{3}{2} v \frac{\partial v}{\partial x_1} + r \frac{\partial^3 v}{\partial x_1^3} &= -s \frac{\partial |u|^2}{\partial x_1},
\end{split}
\end{matrix}
\right.
\label{1975}
\end{equation}
where $k,p,q,r$ and $s$ are real constants, $x_n=\epsilon^n x$, $t_n = \epsilon^n t$. Here $\epsilon$ is the small parameter in terms of which the asymptotic expansions were performed.
This system couples two of the most studied equations in mathematical physics: the KdV equation describes the unidirectional propagation of long, nonlinear dispersive waves, while the cubic nonlinear Schr\"odinger equation governs the slowly varying modulation of a narrow bandwidth train of short waves. Both equations possess important features such as being completely integrable, exhibiting solitary-wave solutions, to name a few \cite{ggkm, sz}. As such, the system (\ref{1975}) is interesting from both mathematical and physical point of view.

However, there are several concerns regarding the above system which have been ignored thus far.  Even though many authors (the papers \cite{albert, pava1, pava2, chen, corcho, dias} are but a small sample of the relevant literature) have studied different mathematical aspects of system (\ref{NLS-KdV}), there exists a tendency to cross reference without checking the details of the original derivation.  Tracing through a plethora of references, the exact derivation of system (\ref{NLS-KdV}) is nowhere to be found. We were led eventually to the paper by Kawahara \textit{et al.} \cite{KSK} which appears to be where the system (\ref{1975}) was first introduced in the context of water waves.  Notice that the first equation in (\ref{1975}) is \textit{linear} whilst that in (\ref{NLS-KdV}) is \textit{nonlinear}. Further, the time scales appearing in (\ref{1975}) are inconsistent, with the dynamics of the second equation of (\ref{1975}) appearing on a slower time scale than that of the first equation. More on this is discussed below. The same is true for the derivation in the context of plasma physics, see \cite{appert, ikezi, nishikawa}, where references lead back to \cite{Zak} and the system (\ref{NLS-KdV}) is not found in any form. Thus it appears that works heretofore studying (\ref{NLS-KdV}) are investigating the mathematical aspects of a \textit{hypothetical} system that has never been derived consistently. Of course, these mathematical considerations are perfectly valid in their own right, but it should be stated that to this point, the results presented are yet to be shown relevant in the context of any application.  Many authors refer to one or multiple of \cite{funakoshi, ikezi, KSK, nishikawa} and each other to motivate the use of the system (\ref{NLS-KdV}), while apparently the details and the results presented in  these papers are ignored.

Even the derivation of system (\ref{1975}) in \cite{KSK} is problematic.  Starting from the Euler water wave problem, the authors introduce multiple spatial and temporal scales $x_n = \epsilon^n x$ and $t_n= \epsilon^n t$ with $x_0=x$ and $t_0=t$ to expand the velocity potential and surface elevation functions in asymptotic series, while assuming that the waves travel in one direction.  At the orders $\epsilon^4$ and $\epsilon^5$, the equations of (\ref{1975}) arise as a consequence of eliminating secular terms. It is immediately clear that the system (\ref{1975}) is troublesome as the two equations appear at different time and spatial scales.  This is dealt with in \cite{KSK} by rewriting the final equations in terms of the first-order slow variables $x_1$ and $t_1$: $t_2=\epsilon t_1$, $t_3=\epsilon^2 t_1$ and $x_2=\epsilon x_1$, $x_3=\epsilon^2 x_1$. Of course this is an inconsistent argument: the different equations encountered to this point are obtained by equating terms at the same order of $\epsilon$. Reintroducing $\epsilon$ at a later point invalidates all calculations to this point.  As pointed out in \cite{DNS}, the main problems with the applicability of (\ref{NLS-KdV}) can be summarized as thus:

\begin{itemize}

\item[(A)] Only a system coupling the \textit{linear} Schr\"odinger (LS) equation with the KdV equation has ever been derived in (\ref{1975}) (see also \cite{funakoshi, ikezi, nishikawa}).

\item[(B)] In the two coupled equations, two \textit{different} time scales appear.

\end{itemize}

Moreover, our calculations and results obtained in \cite{DNS} indicate the impossibility for the derivation of (\ref{NLS-KdV}) in the context of any physical system describing the interaction of long and short waves in dispersive media. It appears impossible even to derive (\ref{1975}) with both equations appearing at the same order.  But those results are still suggestions nevertheless, as the arguments are based on the fifth-order KdV equation which is just an approximation to the Euler equations.  To fully dispute the validity of the system (\ref{1975}), one needs to start the analysis from the full Euler equations.

In this manuscript, we will show that in addition to the aforementioned problems A) and B), indeed neither systems (\ref{NLS-KdV}) nor (\ref{1975}) can be derived consistently from the full Euler equations formulated in the study of water waves--a confirmation to the suggestions raised earlier in \cite{DNS}.

\section{Fundamental equations and summary of results}
In this paper, we consider two-dimensional capillary-gravity waves on an inviscid, incompressible fluid layer of uniform depth $h$ which are governed by the Euler equations
\begin{flalign}
 &\frac{\partial^2 \phi}{\partial x^2}+\frac{\partial^2 \phi}{\partial y^2}=0,\quad    x\in\mathbb{R},\quad
  -1\leq y\leq \eta(x,t);\label{equ: poisson}\\
  &\frac{\partial \eta}{\partial t}- \frac{\partial \phi}{\partial y}+ \frac{\partial \phi}{\partial x} \frac{\partial \eta}{\partial x}=0, \quad \text{at} \  y=\eta(x,t);\label{equ: boundary1}\\
  &\frac{\partial \phi}{\partial t}+\frac{1}{2}\Big\{( \frac{\partial \phi}{\partial x})^2+( \frac{\partial \phi}{\partial y})^2\Big\}+\eta - \frac{1}{W} \frac{\partial^2 \eta}{\partial x^2} \Big\{1+ (\frac{\partial \eta}{\partial x})^2\Big\}^{-3/2}=0,\quad \text{at} \  y=\eta(x,t);\label{equ: boundary2}\\
  &\frac{\partial \phi}{\partial y}=0,\quad \text{at} \ y=-1.\label{equ: boundary3}
\end{flalign}
The harmonic function $\phi(x,y,t)$ is the velocity potential describing the ir-rotational motion of such fluid for $-\infty<x<\infty$, $-1\leq y\leq \eta(x,t)$ and $t$ being the temporal variable.   The $x$-axis is taken along the undisturbed free surface $y=0$, the $y$-axis is taken vertically upwards, and $y=\eta(x,t)$ denotes the elevation of the free surface measured from the undisturbed level.  All the quantities have been normalized by the characteristic length $h$ and the characteristic speed $\sqrt{gh}$, $g$ being the gravitational acceleration.  The Weber number $W$ is defined as $\rho g h^2/T$, where $\rho$ is the fluid density and $T$ the surface tension.  We will make every effort to preserve the usage of notations in \cite{KSK} as we carry out direct comparisons.

Similar to the approach used in \cite{KSK}, we consider the following asymptotic series
\begin{equation*}\label{fun: eta}
  \eta(x,t;\epsilon)=\sum_{n=1}^{\infty}\epsilon^n\eta_n(x_0,x_1,x_2,\cdots,t_0,t_1,t_2,\cdots),
\end{equation*}
and
\begin{equation*}\label{fun: phi}
  \phi(x,y,t;\epsilon)=\sum_{n=1}^{\infty}\epsilon^n\phi_n(x_0,x_1,x_2,\cdots,y,t_0,t_1,t_2,\cdots),
\end{equation*}
with
\begin{equation*}
  x_0=x,\quad t_0=t,\quad x_n=\epsilon^nx,\quad t_n=\epsilon^nt, \ n=1,2,\cdots
\end{equation*}
and expand the equations \eqref{equ: boundary1},\ \eqref{equ: boundary2} around $y=0$.  From these, we obtain a sequence of sets of equations for $\eta_n$ and $\phi_n$ from the coefficients of the like powers in $\epsilon.$  Notice that in \cite{KSK}, the series expansion for $\eta$ started at $n=2$ while that for $\phi$ started at $n=1$.  This is somewhat awkward even with the physical explanation given in there.  Here, we start our expansions for both $\phi$ and $\eta$ at the same point $n=1$ and let the dynamics of the equations \eqref{equ: poisson}-\eqref{equ: boundary3} dictate the relations.  It turns out that to have consistency, $\eta_1 \equiv 0$ and thus all of our calculations agree with those obtained in \cite{KSK} up to the order $\epsilon^3$, except for the presence of the homogeneous solutions which in principle could be omitted, as were taken in \cite{KSK}.  However, at the even orders of $\epsilon^2$ and $\epsilon^4$, the complex amplitudes associated with the non-homogeneous equations arose from \eqref{equ: poisson} must vanish due to certain consistency conditions.  To be precise, and this is the crux of the matter as to why one can neither derive the system (\ref{NLS-KdV}) nor (\ref{1975}),  we discuss this point in detail here (see also the equations \eqref{relation}, \eqref{relation3} and \eqref{relation4} below).  Let $A(x_1,t_1,\ldots)$ be the complex amplitude representing a train of high frequency short waves, $\xi_2(x_1,t_1,\ldots)$ be a slowly varying low frequency long wave and $B(x_1,t_1,\ldots)$ be the homogeneous solution to the harmonic equation associated with \eqref{equ: poisson}.  At the order $\epsilon$, one of the conditions obtained is the dispersion relation
\begin{equation*}
  w^2=(k+\frac{k^3}{W})\tanh k.
\end{equation*}
At the order of $\epsilon^2$, the corresponding equation for \eqref{equ: boundary2} reads
\begin{equation*}
Fe^{i\theta}+c.c.+Ge^{2\theta i}+c.c.+H=0,
\end{equation*}
with c.c. signifies the complex conjugate of the previous term and where the functions $F,G$ and $H$ are given by the following expressions
\begin{equation*}
\begin{split}
 F=&\Big\{\frac{i w^2}{k} - \frac{2ik}{W} - \frac{i}{k}(1+\frac{k^2}{W})(1+k \cosh k)\Big\}\frac{\partial A}{\partial x_1}+\Big\{ -\frac{iw}{k\tanh k} - \frac{i}{w}(1+\frac{k^2}{W})\Big\}\frac{\partial A}{\partial t_1}
  \\
  &+\big(1+\frac{k^2}{W} - \frac{w^2}{k\tanh k}\big) B;
\end{split}
\end{equation*}
\begin{equation*}
  G=\bigg\{\frac{1}{2}\big( \frac{w^2}{\tanh^2k} -w^2\big)+(1+\frac{4k^2}{W})\frac{k}{2\tanh k}\bigg\}A^2;
\end{equation*}
and
\begin{equation*}
  H=\frac{\partial \psi_1}{\partial t_1}+\frac{w^2|A|^2}{\tanh^2k} +w^2|A|^2+\xi_2.
\end{equation*}
Thus, it must be the case that $F=G=H=0$.  In particular, the complex amplitude function $A$ must vanish in order that the dispersion relation mentioned above holds true.
The inclusion of homogeneous solutions as mentioned above therefore is necessary.  Similar expression appears at the order of $\epsilon^4$ for the corresponding equation for
\eqref{equ: boundary2}.  The vanishing condition for $G$ is in direct contradiction with the dispersion relation unless $A\equiv 0$.  This is the reason why one \textbf{cannot} derive the couple NLS-KdV system (\ref{NLS-KdV}) from the context of the Euler equations.  In principle it could be possible to alter some terms in the Euler equations \eqref{equ: poisson}-\eqref{equ: boundary3} so that second harmonic resonance occurs (that is, the coefficient for $G$ vanishes) but then we would no longer be dealing with the Euler water wave problem mentioned above.

Starting from the order $\epsilon^4$, our results deviate significantly from those obtained in \cite{KSK}.  In particular, the linear Schr\"odinger equation in the system
(\ref{1975}), \textit{i.e.}, the first equation, appears at the \textit{fifth} order instead of \textit{fourth} in our results.  Moreover, at this order $\epsilon^5$ we also
obtain the \textit{pure} KdV-equation \textit{without any coupling term}.  As none of the calculations were presented after the order of $\epsilon^3$ in the paper \cite{KSK}, we cannot finger point precisely at what point things start to differ and/or what assumptions had been made to rule out the vanishing of $A$ in that paper.

Our analysis establishes unequivocally the fact that, contrary to what has been assumed heretofore, one \textbf{can neither} derive the couple linear Schr\"odinger-KdV system (\ref{1975}) \textbf{nor} the NLS-KdV system (\ref{NLS-KdV}) from the Euler equations used in the study of water waves.
\section{The main results}
\subsection{At the order $O(\epsilon)$}
The first order problem is as follows:
\begin{flalign}
&\frac{\partial^2 \phi_1}{\partial x_0^2}+\frac{\partial^2 \phi_1}{\partial y^2}=0;\label{equ: poisson1}\\
& \frac{\partial \eta_1}{\partial t_0}- \frac{\partial \phi_1}{\partial y}=0, \quad \text{at} \  y=0;\label{equ: boundaryI-1}\\
&\frac{\partial \phi_1}{\partial t_0}+\eta_1-\frac{1}{W}\frac{\partial^2 \eta_1}{\partial x_0^2}=0,\quad \text{at} \  y=0;\label{equ: boundaryII-1}\\
& \frac{\partial \phi_1}{\partial y}=0,\quad \text{at} \ y=-1.\label{equ: boundaryIII-1}
\end{flalign}
We take as our starting point the solution for the harmonic function $\phi_1$ as the superposition of the short and long waves of the following form
\begin{equation}\label{functions:phi}
  \phi_1=f(y)A(x_1,t_1,\cdots)e^{i\theta}+c.c.+\psi_1(x_1,t_1,\cdots)
\end{equation}
where $\theta=kx_0-wt_0$ and $A\in \mathbb{C},\ \ \psi_1\in \mathbb{R}$ are functions independent of the slow variables $x_0,t_0,y$. Substituting \eqref{functions:phi} into \eqref{equ: poisson1}, we have
\begin{equation*}
  A[f''(y)-k^2f(y)]e^{i\theta}+c.c.=0.
\end{equation*}
Thus, either $A=0$ or
\begin{equation}\label{ODE1}
  f''(y)-k^2f(y)=0.
\end{equation}
The ordinary differential equation \eqref{ODE1} has the solution
\begin{equation*}
  f(y)=C_1e^{ky}+C_2e^{-ky},
\end{equation*}
while the boundary condition \eqref{equ: boundaryIII-1} implies that
\begin{equation*}
  kC_1e^{-k}-kC_2e^{k}=0.
\end{equation*}
Consequently, it is deduced that
\begin{equation*}
  \phi_1=C\cosh [k(y+1)]A(x_1,t_1,\cdots)e^{i\theta}+c.c.+\psi_1(x_1,t_1,\cdots).
\end{equation*}
Next, it follows from equation \eqref{equ: boundaryI-1} that
\begin{equation*}
  \eta_1=\frac{Ck\sinh k}{-iw}Ae^{i\theta}+c.c.+\xi_1(x_1,t_1,\cdots)
\end{equation*}
for some real function $\xi_1(x_1,t_1,\cdots).$   In order to make a direct comparison with the paper \cite{KSK}, we let $C=\frac{-iw}{k\sinh k}$ and obtain
\begin{equation*}\label{phi1}
 \phi_1=\frac{-iw}{k\sinh k}\cosh [k(y+1)]Ae^{i\theta}+c.c.+\psi_1(x_1,t_1,\cdots),
\end{equation*}
and
\begin{equation}\label{eta1}
  \eta_1=Ae^{i\theta}+c.c.+\xi_1(x_1,t_1,\cdots).
\end{equation}
Now, putting the expressions for $\phi_1$ and $\eta_1$ obtained above into \eqref{equ: boundaryII-1}, it reveals that
\begin{equation}\label{dispersion}
  \frac{-w^2\cosh k}{k\sinh k} Ae^{i\theta}+(1+
  \frac{k^2}{W})Ae^{i\theta}+c.c.+\xi_1=0.
\end{equation}
Noting that $\xi_1$ and $A$ are functions independent of $x_0,t_0$, hence \eqref{dispersion} implies that
\begin{equation*}
  \xi_1=0,
\end{equation*}
and
\begin{equation*}
  A\Big\{\frac{-w^2\cosh k}{k\sinh k}+(1+
  \frac{k^2}{W})\Big\}=0.
\end{equation*}
In other words, one has the following dispersion relation
\begin{equation}\label{relation}
  w^2=(k+\frac{k^3}{W})\tanh k
\end{equation}
that must hold true unless $A=0.$
It is therefore concluded that the first order problem in $O(\epsilon)$ has solution
\begin{equation*}\label{solution phi1}
  \phi_1=\frac{-iw}{k\sinh k}\cosh [k(y+1)]Ae^{i\theta}+c.c.+\psi_1,
\end{equation*}
and
\begin{equation*}
  \eta_1=Ae^{i\theta}+c.c.,
\end{equation*}
along with the dispersion relation \eqref{relation}, where $A$ and $\psi_1$ are two arbitrary functions with respect to the variables $x_1,t_1,x_2,t_2,\cdots$ but independent of the slow variables $x_0,t_0,y$.
\subsection{At the order $O(\epsilon^2)$}
The second order problem is as follows:
\begin{flalign}
&\frac{\partial^2 \phi_2}{\partial x_0^2}+\frac{\partial^2 \phi_2}{\partial y^2}=-2\frac{\partial^2\phi_1}{\partial x_0\partial x_1};\label{equ: poisson2}\\
&\frac{\partial \eta_2}{\partial t_0}- \frac{\partial \phi_2}{\partial y}=-\frac{\partial \eta_1}{\partial t_1}-\frac{\partial \phi_1}{\partial x_0}\frac{\partial \eta_1}{\partial x_0}, \quad \text{at} \  y=0;\label{equ: boundaryI-2}\\
&\frac{\partial \phi_2}{\partial t_0}+\frac{\partial \phi_1}{\partial t_1}+\frac{1}{2}\Big\{( \frac{\partial \phi_1}{\partial x_0})^2+( \frac{\partial \phi_1}{\partial y})^2\Big\}+\eta_2-\frac{1}{W}(\frac{\partial^2 \eta_2}{\partial x_0^2}+2\frac{\partial^2 \eta_1}{\partial x_0\partial x_1})=0,\quad \text{at} \  y=0;\label{equ: boundaryII-2}\\
&\frac{\partial \phi_2}{\partial y}=0,\quad \text{at} \ y=-1.\label{equ: boundaryIII-2}
\end{flalign}
Putting the form of $\phi_1$ obtained from the order $O(\epsilon)$ above into \eqref{equ: poisson2} we have
\begin{equation*}
  \frac{\partial^2 \phi_2}{\partial x_0^2}+\frac{\partial^2 \phi_2}{\partial y^2}=-\frac{2w}{\sinh k}\cosh [k(y+1)]\frac{\partial A}{\partial x_1}e^{i\theta}+c.c.,
\end{equation*}
which has solution
\begin{equation}\label{solu:phi2}
  \phi_2=-\frac{w(y+1)}{k\sinh k}\sinh [k(y+1)]\frac{\partial A}{\partial x_1}e^{i\theta}+\frac{-iw}{k\sinh k}\cosh [k(y+1)]Be^{i\theta}+c.c.+\psi_2,
\end{equation}
where $B$ and $\psi_2$ are arbitrary functions independent of the slow variables $x_0,t_0,y$.
Next, it follows from
\eqref{equ: boundaryI-2} that
\begin{equation*}
\begin{split}
  \frac{\partial \eta_2}{\partial t_0}&= \frac{\partial \phi_2}{\partial y}-\frac{\partial A}{\partial t_1}e^{i\theta}-\frac{ikw}{\tanh k}A^2e^{2\theta i}+c.c.\\
  &=-\frac{w}{k}(1+k\coth k)\frac{\partial A}{\partial x_1}e^{i\theta}-iwBe^{i\theta}-\frac{\partial A}{\partial t_1}e^{i\theta}-\frac{ikw}{\tanh k}A^2e^{2\theta i}+c.c..
  \end{split}
\end{equation*}
Integrating the above equation with respect to  $t_0$ we obtain
\begin{equation}\label{solu:eta2}
  \eta_2=-\frac{i}{k}(1+k\coth k)\frac{\partial A}{\partial x_1}e^{i\theta}+Be^{i\theta}-\frac{i}{w}\frac{\partial A}{\partial t_1}e^{i\theta}+\frac{k}{2\tanh k}A^2e^{2\theta i}+c.c.+\xi_2,
\end{equation}
where $\xi_2$ is an arbitrary function independent of the slow variables $x_0,t_0,y.$
Substituting \eqref{solu:phi2} and \eqref{solu:eta2} into \eqref{equ: boundaryII-2}, we deduce that
\begin{equation*}\label{relation2}
\begin{split}
&\frac{iw^2}{k}\frac{\partial A}{\partial x_1}e^{i\theta}+\frac{-w^2}{k\tanh k} Be^{i\theta}+\frac{-iw}{k\tanh k}\frac{\partial A}{\partial t_1}e^{i\theta}+\frac{\partial \psi_1}{\partial t_1}\\
&+\frac{1}{2}\Big\{ \frac{w^2A^2}{\tanh^2k}e^{2\theta i}+2\frac{w^2|A|^2}{\tanh^2k} -w^2A^2e^{2\theta i}+2w^2|A|^2\Big\}\\
&+(1+\frac{k^2}{W})\Big\{-\frac{i}{k}(1+k\coth k)\frac{\partial A}{\partial x_1}e^{i\theta}+Be^{i\theta}-\frac{i}{w}\frac{\partial A}{\partial t_1}e^{i\theta}\Big\}\\
&+(1+\frac{4k^2}{W})\frac{k}{2\tanh k}A^2e^{2\theta i}+\xi_2-\frac{2ik}{W}\frac{\partial A}{\partial x_1}e^{i\theta}+c.c.=0.\\
\end{split}
\end{equation*}
The above equation can be rewritten in the following compact form
\begin{equation}\label{relation3}
F e^{i\theta}+G e^{2\theta i}+c.c.+H=0,
\end{equation}
where the functions $F,G$ and $H$ are given by the following expressions
\begin{equation*}
\begin{split}
  F&=\frac{iw^2}{k}\frac{\partial A}{\partial x_1}+\frac{-iw}{k\tanh k}\frac{\partial A}{\partial t_1}
  +(1+\frac{k^2}{W})\Big\{-\frac{i}{k}(1+k\coth k)\frac{\partial A}{\partial x_1}-\frac{i}{w}\frac{\partial A}{\partial t_1}\Big\}\\&\quad-\frac{2ik}{W}\frac{\partial A}{\partial x_1}+\frac{-w^2}{k\tanh k} B+(1+\frac{k^2}{W})B;
  \end{split}
\end{equation*}
\begin{equation*}
  G=\frac{1}{2}\Big\{ \frac{w^2A^2}{\tanh^2k} -w^2A^2\Big\}+(1+\frac{4k^2}{W})\frac{k}{2\tanh k}A^2;
\end{equation*}
and
\begin{equation*}
  H=\frac{\partial \psi_1}{\partial t_1}+\frac{w^2|A|^2}{\tanh^2k} +w^2|A|^2+\xi_2.
\end{equation*}
Thus, \eqref{relation3} implies that
\begin{equation}\label{relation4}
  F=G=H=0.
\end{equation}
In particular, $G=0$ implies that $A=0$, otherwise a contradiction to the dispersion relation \eqref{dispersion} will arise.   Thus, \eqref{relation4} reduces to the condition
\begin{equation}\label{relation7}
\frac{\partial \psi_1}{\partial t_1}+\xi_2=0,
\end{equation}
apart from \eqref{relation}.
We therefore conclude that
\begin{equation*}
  \eta_1=0,\ \ \phi_1=\psi_1,
\end{equation*}
\begin{equation}\label{solution phi2}
  \phi_2=\frac{-iw}{k\sinh k}\cosh [k(y+1)]Ae^{i\theta}+c.c.+\psi_2,
\end{equation}
and
\begin{equation}\label{solution eta2}
  \eta_2=Ae^{i\theta}+c.c.+\xi_2,
\end{equation}
where, in the interest of preserving the same notations as in \cite{KSK} for direct comparison, \textbf{we have renamed $B$ by $A.$}\\

\noindent \textbf{Remark 1.} The second term with the function $B$ in the right hand side of \eqref{solu:phi2}, which represents the homogeneous solution,  was omitted in \cite{KSK}.  It is now clear that its inclusion is necessary. Notice that in \cite{KSK} the asymptotic expansion for $\phi$ started at $n=1$ while that for $\eta$ started at $n=2$.  This is somewhat awkward.  Here, we start both of the expansions at the same point $n=1$ and let the dynamics of the problem dictate the relation.  In order to have consistency, $\eta_1$ must vanish and we arrive at the exact same results as obtained in \cite{KSK}.

\subsection{At the third order in $O(\epsilon^3)$}
The third order problem is as follows:
\begin{flalign}
&\frac{\partial^2 \phi_3}{\partial x_0^2}+\frac{\partial^2 \phi_3}{\partial y^2}=-2\frac{\partial^2\phi_2}{\partial x_0\partial x_1}-\frac{\partial^2\phi_1}{\partial^2 x_1};\label{equ: poisson3}\\
&\frac{\partial \eta_3}{\partial t_0}- \frac{\partial \phi_3}{\partial y}=-\frac{\partial \eta_2}{\partial t_1}, \quad \text{at} \  y=0;\label{equ: boundaryI-3}\\
&\frac{\partial \phi_3}{\partial t_0}+\frac{\partial \phi_2}{\partial t_1}+\frac{\partial \phi_1}{\partial t_2}+\eta_3-\frac{1}{W}(\frac{\partial^2 \eta_3}{\partial x_0^2}+2\frac{\partial^2 \eta_2}{\partial x_0\partial x_1})=0,\quad \text{at} \  y=0;\label{equ: boundaryII-3}\\
&\frac{\partial \phi_3}{\partial y}=0,\quad \text{at} \ y=-1.\label{equ: boundaryIII-3}
\end{flalign}
Introducing the form of $\phi_2$ obtained from \eqref{solution phi2} into \eqref{equ: poisson3}, we have
\begin{equation*}
  \frac{\partial^2 \phi_3}{\partial x_0^2}+\frac{\partial^2 \phi_3}{\partial y^2}=-\frac{2w}{\sinh k}\cosh [k(y+1)]\frac{\partial A}{\partial x_1}e^{i\theta}+c.c-\frac{\partial^2\psi_1}{\partial x_1^2}.
\end{equation*}
This has a solution given by the following form
\begin{equation}\label{solu:phi3}
  \phi_3=-\frac{w(y+1)}{k\sinh k}\sinh [k(y+1)]\frac{\partial A}{\partial x_1}e^{i\theta}+\frac{-iw}{k\sinh k}\cosh [k(y+1)]Be^{i\theta}+c.c.-\frac{1}{2}\frac{\partial^2\psi_1}{\partial x_1^2}(y+1)^2+\psi_3,
\end{equation}
where $B\in \mathbb{C}$ (the homogeneous solution) and $\psi_3\in\mathbb{R}$ are arbitrary functions independent of the slow variables $x_0,t_0,y$.  Next, it follows from
\eqref{equ: boundaryI-3} that
\begin{equation*}
\begin{split}
  \frac{\partial \eta_3}{\partial t_0}&= \frac{\partial \phi_3}{\partial y}-\frac{\partial A}{\partial t_1}e^{i\theta}+c.c.-\frac{\partial \xi_2}{\partial t_1}\\
  &=-\frac{w}{k}(1+k\coth k)\frac{\partial A}{\partial x_1}e^{i\theta}-iwBe^{i\theta}-\frac{\partial^2\psi_1}{\partial x_1^2}-\frac{\partial A}{\partial t_1}e^{i\theta}+c.c.-\frac{\partial \xi_2}{\partial t_1}.
  \end{split}
\end{equation*}
Integrating the above equation with respect to  $t_0$ we obtain
\begin{equation}\label{solu:eta3}
  \eta_3=-\frac{i}{k}(1+k\coth k)\frac{\partial A}{\partial x_1}e^{i\theta}+Be^{i\theta}-\frac{i}{w}\frac{\partial A}{\partial t_1}e^{i\theta}+c.c.+\xi_3,
\end{equation}
where $\xi_3\in\mathbb{R}$ is an arbitrary function independent of the slow variables $x_0,t_0,y$, provided that the non-secularity condition
\begin{equation}\label{relation8}
 \frac{\partial \xi_2}{\partial t_1}+\frac{\partial^2\psi_1}{\partial x_1^2}=0
\end{equation}
is satisfied.  Substituting \eqref{solu:phi3} and \eqref{solu:eta3} into \eqref{equ: boundaryII-3}, we arrive at
\begin{equation*}\label{relation5}
\begin{split}
&\frac{iw^2}{k}\frac{\partial A}{\partial x_1}e^{i\theta}-\frac{w^2}{k\tanh k} Be^{i\theta}-\frac{iw}{k\tanh k}\frac{\partial A}{\partial t_1}e^{i\theta}+\frac{\partial \psi_2}{\partial t_1}+\frac{\partial \psi_1}{\partial t_2}\\
&+(1+\frac{k^2}{W})\Big\{-\frac{i}{k}(1+k\coth k)\frac{\partial A}{\partial x_1}e^{i\theta}+Be^{i\theta}-\frac{i}{w}\frac{\partial A}{\partial t_1}e^{i\theta}\Big\}\\
&+\xi_3-\frac{2ik}{W}\frac{\partial A}{\partial x_1}e^{i\theta}+c.c.=0.\\
\end{split}
\end{equation*}
From this, one can see that
\begin{equation}\label{relation6}
  \frac{iw^2}{k}\frac{\partial A}{\partial x_1}-\frac{iw}{k\tanh k}\frac{\partial A}{\partial t_1}+(1+\frac{k^2}{W})\Big\{-\frac{i}{k}(1+k\coth k)\frac{\partial A}{\partial x_1}-\frac{i}{w}\frac{\partial A}{\partial t_1}\Big\}-\frac{2ik}{W}\frac{\partial A}{\partial x_1}=0,
\end{equation}
and
\begin{equation}\label{relation9}
\frac{\partial \psi_2}{\partial t_1}+\frac{\partial \psi_1}{\partial t_2}+\xi_3=0,
\end{equation}
as the coefficient for $B$ happens to satisfy the dispersive relation \eqref{relation}.

The equation \eqref{relation6} can be rewritten as
\begin{equation}\label{non-seculartiy2}
\frac{\partial A}{\partial t_1}+V_g\frac{\partial A}{\partial x_1}=0,
\end{equation}
where $V_g$ denotes the group velocity
\begin{equation}
  V_g=\frac{dw}{dk}=\frac{1}{2w}\Big[(1+\frac{3k^2}{W})\tanh k+(k+\frac{k^3}{W}){\rm sech}^2 k\Big].
\end{equation}
Eliminating of $\xi_2$ between \eqref{relation7} and \eqref{relation8} yields
\begin{equation}\label{*}
  \frac{\partial^2 \psi_1}{\partial t_1^2}-\frac{\partial^2 \psi_1}{\partial x_1^2}=0.
\end{equation}

\noindent\textbf{Remark 2.} Thus, up to the order of $O(\epsilon^3)$, all of our analysis reveals exact same results as those obtained in \cite{KSK}.  Starting from the next order, however, the analysis will differ significantly.

\subsection{At the fourth order in $O(\epsilon^4)$}
The fourth order problem is as follows:
\begin{flalign}
&\frac{\partial^2 \phi_4}{\partial x_0^2}+\frac{\partial^2 \phi_4}{\partial y^2}=-2\frac{\partial^2\phi_3}{\partial x_0\partial x_1}-2\frac{\partial^2\phi_2}{\partial x_0\partial x_2}-\frac{\partial^2\phi_2}{\partial x_1^2}-2\frac{\partial^2\phi_1}{\partial x_1\partial x_2};\label{equ: poisson4}\\
&\frac{\partial \eta_4}{\partial t_0}- \frac{\partial \phi_4}{\partial y}=-\frac{\partial \eta_3}{\partial t_1}-\frac{\partial \eta_2}{\partial t_2}-\frac{\partial \phi_1}{\partial x_1}\frac{\partial \eta_2}{\partial x_0}-\frac{\partial \phi_2}{\partial x_0}\frac{\partial \eta_2}{\partial x_0}, \quad \text{at} \  y=0;\label{equ: boundaryI-4}\\
&\frac{\partial \phi_4}{\partial t_0}+\frac{\partial \phi_3}{\partial t_1}+\frac{\partial \phi_2}{\partial t_2}+\frac{\partial \phi_1}{\partial t_3}+\frac{1}{2}\Big\{(\frac{\partial \phi_1}{\partial x_1}+\frac{\partial \phi_2}{\partial x_0})^2+(\frac{\partial \phi_2}{\partial y})^2\Big\}\nonumber\\
&\quad+\eta_4-\frac{1}{W}(\frac{\partial^2 \eta_4}{\partial x_0^2}+2\frac{\partial^2 \eta_3}{\partial x_0\partial x_1}+2\frac{\partial^2 \eta_2}{\partial x_0\partial x_2}+\frac{\partial^2 \eta_2}{\partial x_1^2})=0,\quad \text{at} \  y=0;\label{equ: boundaryII-4}\\
&\frac{\partial \phi_4}{\partial y}=0,\quad \text{at} \ y=-1.\label{equ: boundaryIII-4}
\end{flalign}
Introducing \eqref{solution phi2} and \eqref{solu:phi3} into \eqref{equ: poisson4}, we have
\begin{equation*}
\begin{split}
  \frac{\partial^2 \phi_4}{\partial x_0^2}+\frac{\partial^2 \phi_4}{\partial y^2}&=\frac{2iw(y+1)}{\sinh k}\sinh [k(y+1)]\frac{\partial^2 A}{\partial x_1^2}e^{i\theta} -\frac{2w}{\sinh k}\cosh [k(y+1)]\frac{\partial B}{\partial x_1}e^{i\theta}\\
  &-\frac{2w}{\sinh k}\cosh [k(y+1)]\frac{\partial A}{\partial x_2}e^{i\theta} +\frac{iw}{k\sinh k}\cosh [k(y+1)]\frac{\partial^2 A}{\partial x_1^2}e^{i\theta}+c.c.-\frac{\partial^2 \psi_2}{\partial x_1^2}-2\frac{\partial^2 \psi_1}{\partial x_1\partial x_2}.\\
  \end{split}
\end{equation*}
This equation has a solution given by the following form
\begin{equation}\label{solu:phi4}
\begin{split}
  \phi_4&=-\frac{w(y+1)}{k\sinh k}\sinh [k(y+1)]\frac{\partial A}{\partial x_2}e^{i\theta}+\frac{iw(y+1)^2}{2k\sinh k}\cosh [k(y+1)]\frac{\partial^2 A}{\partial x_1^2}e^{i\theta}\\
  &-\frac{w(y+1)}{k\sinh k}\sinh [k(y+1)]\frac{\partial B}{\partial x_1}e^{i\theta}+\frac{-iw}{k\sinh k}\cosh [k(y+1)]\widetilde{B}e^{i\theta}+c.c.\\
  &-\frac{1}{2}\frac{\partial^2 \psi_2}{\partial x_1^2}(y+1)^2-\frac{\partial^2 \psi_1}{\partial x_1\partial x_2}(y+1)^2+\psi_4,\\
  \end{split}
\end{equation}
where $\widetilde{B}\in \mathbb{C}$ (the homogeneous solution) and $\psi_4\in\mathbb{R}$ are arbitrary functions independent of the slow variables $x_0,t_0,y$.  Next, by putting the expressions previously obtained for $\phi_4,\eta_3,\eta_2$ into \eqref{equ: boundaryI-4}, we have
\begin{equation*}
\begin{split}
  \frac{\partial \eta_4}{\partial t_0}&= \frac{\partial \phi_4}{\partial y}-\frac{\partial \eta_3}{\partial t_1}-\frac{\partial \eta_2}{\partial t_2}-\frac{\partial \phi_1}{\partial x_1}\frac{\partial \eta_2}{\partial x_0}-\frac{\partial \phi_2}{\partial x_0}\frac{\partial \eta_2}{\partial x_0}\\
  &=\Big\{-\frac{w}{k}(1+k\coth k)\frac{\partial A}{\partial x_2}e^{i\theta}+\frac{iw}{2k}(k+2\coth k)\frac{\partial^2 A}{\partial x_1^2}e^{i\theta}\\
  &\quad-\frac{w}{k}(1+k\coth k)\frac{\partial B}{\partial x_1}e^{i\theta}-iw\widetilde{B}e^{i\theta}
  -\frac{\partial^2 \psi_2}{\partial x_1^2}-2\frac{\partial^2 \psi_1}{\partial x_1\partial x_2}\Big\}\\
  &-\Big\{-\frac{i}{k}(1+k\coth k)\frac{\partial^2 A}{\partial x_1\partial t_1}e^{i\theta}+\frac{\partial B}{\partial t_1}e^{i\theta}-\frac{i}{w}\frac{\partial^2 A}{\partial t_1^2}e^{i\theta}+\frac{\partial \xi_3}{\partial t_1}\Big\}\\
  &-\Big\{\frac{\partial A}{\partial t_2}e^{i\theta}+\frac{\partial \xi_2}{\partial t_2}\Big\}-ik\frac{\partial \phi_1}{\partial x_1}Ae^{i\theta}-\frac{ikw}{\tanh k}A^2e^{i2\theta}+c.c..\\
\end{split}
\end{equation*}
An integration with respect to the slow variable $t_0$ gives
\begin{equation*}
\begin{split}
\eta_4
  &=\Big\{-\frac{i}{k}(1+k\coth k)\frac{\partial A}{\partial x_2}e^{i\theta}-\frac{1}{2k}(k+2\coth k)\frac{\partial^2 A}{\partial x_1^2}e^{i\theta}\\
  &\quad-\frac{i}{k}(1+k\coth k)\frac{\partial B}{\partial x_1}e^{i\theta}+\widetilde{B}e^{i\theta}
  \Big\}\\
  &-\Big\{\frac{1}{wk}(1+k\coth k)\frac{\partial^2 A}{\partial x_1\partial t_1}e^{i\theta}+\frac{i}{w}\frac{\partial B}{\partial t_1}e^{i\theta}+\frac{1}{w^2}\frac{\partial^2 A}{\partial t_1^2}e^{i\theta}\Big\}\\
  &-\frac{i}{w}\frac{\partial A}{\partial t_2}e^{i\theta}+\frac{k}{w}\frac{\partial \phi_1}{\partial x_1}Ae^{i\theta}+\frac{k}{2\tanh k}A^2e^{i2\theta}+c.c.+\xi_4,\\
\end{split}
\end{equation*}
provided that the non-secularity condition
\begin{equation}\label{non-seculartiy}
  \frac{\partial^2 \psi_2}{\partial x_1^2}+2\frac{\partial^2 \psi_1}{\partial x_1\partial x_2}
  +\frac{\partial \xi_3}{\partial t_1}+\frac{\partial \xi_2}{\partial t_2}=0
\end{equation}
holds true.  Eliminating of $\xi_2,\xi_3$ between \eqref{relation7}, \eqref{relation9} and \eqref{non-seculartiy} yields
\begin{equation*}
  \frac{\partial^2 \psi_2}{\partial t_1^2}-\frac{\partial^2 \psi_2}{\partial x_1^2}+2(\frac{\partial^2 \psi_1}{\partial t_1\partial t_2}-\frac{\partial^2 \psi_1}{\partial x_1\partial x_2})=0.
\end{equation*}
Introducing $\phi_4$ and $\eta_4$ into \eqref{equ: boundaryII-4}, we then have
\begin{equation}\label{relation11}
\begin{split}
&\frac{\partial \phi_4}{\partial t_0}+\frac{\partial \phi_3}{\partial t_1}+\frac{\partial \phi_2}{\partial t_2}+\frac{\partial \phi_1}{\partial t_3}+\frac{1}{2}\Big\{(\frac{\partial \phi_1}{\partial x_1}+\frac{\partial \phi_2}{\partial x_0})^2+(\frac{\partial \phi_2}{\partial y})^2\Big\}\\
&+\eta_4-\frac{1}{W}(\frac{\partial^2 \eta_4}{\partial x_0^2}+2\frac{\partial^2 \eta_3}{\partial x_0\partial x_1}+2\frac{\partial^2 \eta_2}{\partial x_0\partial x_2}+\frac{\partial^2 \eta_2}{\partial x_1^2})\\
  &=\Big\{\frac{iw^2}{k}\frac{\partial A}{\partial x_2}e^{i\theta}+\frac{w^2}{2k\tanh k}\frac{\partial^2 A}{\partial x_1^2}e^{i\theta}
  +\frac{iw^2}{k}\frac{\partial B}{\partial x_1}e^{i\theta}+\frac{-w^2}{k\tanh k}\widetilde{B}e^{i\theta}\Big\}\\
  &+\Big\{-\frac{w}{k}\frac{\partial^2 A}{\partial x_1\partial t_1}e^{i\theta}+\frac{-iw}{k\tanh k}\frac{\partial B}{\partial t_1}e^{i\theta}-\frac{1}{2}\frac{\partial^3\psi_1}{\partial x_1^2\partial t_1}+\frac{\partial \psi_3}{\partial t_1}\Big\}\\
  &+\frac{-iw}{k\tanh k}\frac{\partial A}{\partial t_2}e^{i\theta}+\frac{\partial \psi_2}{\partial t_2}+\frac{\partial \psi_1}{\partial t_3}\\
  &+\Big\{\frac{1}{2}(\frac{\partial \psi_1}{\partial x_1})^2+\frac{\partial \psi_1}{\partial x_1}\frac{w}{\tanh k}Ae^{i\theta}+\frac{w^2}{2\tanh^2 k} A^2e^{2\theta i}-\frac{1}{2}w^2A^2e^{i2\theta}+\frac{w^2}{\tanh^2 k}|A|^2+w^2|A|^2\Big\}\\
  &+\eta_4-\frac{1}{W}(\frac{\partial^2 \eta_4}{\partial x_0^2}+2\frac{\partial^2 \eta_3}{\partial x_0\partial x_1}+2\frac{\partial^2 \eta_2}{\partial x_0\partial x_2}+\frac{\partial^2 \eta_2}{\partial x_1^2}) + c.c.=0.\\
  \end{split}
\end{equation}
This expression can be written in the compact form of \eqref{relation3} from which the conclusion of $A=0$ must again be drawn.  Consequently, $\phi_2,\eta_2, \phi_3,\eta_3,\phi_4,\eta_4$ are now reduced to
\begin{equation*}
  \phi_2=\psi_2,\quad \quad \quad \eta_2=\xi_2;
\end{equation*}
\begin{equation*}
\phi_3=\frac{-iw}{k\sinh k}\cosh [k(y+1)]Ae^{i\theta}+c.c.-\frac{1}{2}\frac{\partial^2 \psi_1}{\partial x_1^2}(y+1)^2+\psi_3,\quad\quad \quad \eta_3=Ae^{i\theta}+c.c.+\xi_3;
\end{equation*}
and
\begin{equation*}
\begin{split}
  \phi_4&=-\frac{w(y+1)}{k\sinh k}\sinh [k(y+1)]\frac{\partial A}{\partial x_1}e^{i\theta}+\frac{-iw}{k\sinh k}\cosh [k(y+1)]Be^{i\theta}+c.c.\\
  &\quad
  -\frac{1}{2}\frac{\partial^2 \psi_2}{\partial x_1^2}(y+1)^2-\frac{\partial^2 \psi_1}{\partial x_1\partial x_2}(y+1)^2+\psi_4;
  \end{split}
\end{equation*}
\begin{equation*}
  \eta_4= -\frac{i}{k}(1+k\coth k)\frac{\partial A}{\partial x_1}e^{i\theta}+Be^{i\theta}
-\frac{i}{w}\frac{\partial A}{\partial t_1}e^{i\theta}+c.c.+\xi_4;
\end{equation*}
where again, in the interest of preserving the same notations as in \cite{KSK}, \textbf{we use $A,B$ to replace $B,\widetilde{B}$ respectively}.
Thus, \eqref{relation11} reduces to
\begin{equation*}\label{relation12}
\begin{split}
&\Big\{
  \frac{iw^2}{k}\frac{\partial A}{\partial x_1}e^{i\theta}+\frac{-w^2}{k\tanh k}Be^{i\theta}\Big\}
  +\Big\{\frac{-iw}{k\tanh k}\frac{\partial A}{\partial t_1}e^{i\theta}-\frac{1}{2}\frac{\partial^3\psi_1}{\partial x_1^2\partial t_1}+\frac{\partial \psi_3}{\partial t_1}\Big\}\\
  &+\frac{\partial \psi_2}{\partial t_2}+\frac{\partial \psi_1}{\partial t_3}+\frac{1}{2}(\frac{\partial \psi_1}{\partial x_1})^2 +\xi_4-\frac{2ik}{W}\frac{\partial A}{\partial x_1}e^{i\theta}-\frac{1}{W}\frac{\partial^2 \xi_2}{\partial x_1^2}\\
  &+(1+\frac{k^2}{W})\Big\{-\frac{i}{k}(1+k\coth k)\frac{\partial A}{\partial x_1}e^{i\theta}+Be^{i\theta}
-\frac{i}{w}\frac{\partial A}{\partial t_1}e^{i\theta}\Big\} + c.c.=0,\\
  \end{split}
\end{equation*}
from which the following relation
\begin{equation*}
  -\frac{1}{2}\frac{\partial^3\psi_1}{\partial x_1^2\partial t_1}+\frac{\partial \psi_3}{\partial t_1}+\frac{\partial \psi_2}{\partial t_2}+\frac{\partial \psi_1}{\partial t_3}+\frac{1}{2}(\frac{\partial \psi_1}{\partial x_1})^2+\xi_4-\frac{1}{W}\frac{\partial^2\xi_2}{\partial x_1^2}=0,
\end{equation*}
along with \eqref{non-seculartiy2} again are derived.\\

\noindent\textbf{Remark 3.} Because of the vanishing condition \eqref{relation11} on $A$, the linear Schr\"odinger equation ( \textit{i.e.} the first equation in \eqref{1975}) does not appear in our analysis at this order of $\epsilon^4$ as in \cite{KSK}.  Since none of the calculations were  presented after the order of $\epsilon^3$ in the paper \cite{KSK}, we cannot finger point precisely at what point things start to differ and/or what assumptions had been made to rule out the vanishing of $A$ in that paper.

\subsection{At the fifth order in $O(\epsilon^5)$}
The fifth order problem is as follows:
\begin{flalign}
&\frac{\partial^2 \phi_5}{\partial x_0^2}+\frac{\partial^2 \phi_5}{\partial y^2}=-2\frac{\partial^2\phi_4}{\partial x_0\partial x_1}-2\frac{\partial^2\phi_3}{\partial x_0\partial x_2}-\frac{\partial^2\phi_3}{\partial x_1^2}-2\frac{\partial^2\phi_2}{\partial x_1\partial x_2}-\frac{\partial^2\phi_1}{\partial x_2^2}-2\frac{\partial^2\phi_1}{\partial x_1\partial x_3};\label{equ: poisson5}\\
&\frac{\partial \eta_5}{\partial t_0}- \frac{\partial \phi_5}{\partial y}=-\frac{\partial \eta_4}{\partial t_1}-\frac{\partial \eta_3}{\partial t_2}-\frac{\partial \eta_2}{\partial t_3}-\frac{\partial \phi_1}{\partial x_1}\frac{\partial \eta_2}{\partial x_1}, \quad \text{at} \  y=0;\label{equ: boundaryI-5}\\
&\frac{\partial \phi_5}{\partial t_0}+\frac{\partial \phi_4}{\partial t_1}+\frac{\partial \phi_3}{\partial t_2}+\frac{\partial \phi_2}{\partial t_3}+\frac{\partial \phi_1}{\partial t_4}+\frac{\partial \phi_1}{\partial x_1}\Big[\frac{\partial \phi_1}{\partial x_2}+\frac{\partial \phi_2}{\partial x_1}+\frac{\partial \phi_3}{\partial x_0}\Big]\nonumber\\
&\quad+\eta_5-\frac{1}{W}\Big(\frac{\partial^2 \eta_5}{\partial x_0^2}+2\frac{\partial^2 \eta_4}{\partial x_0\partial x_1}+2\frac{\partial^2 \eta_3}{\partial x_0\partial x_2}+\frac{\partial^2 \eta_3}{\partial x_1^2}+2\frac{\partial^2 \eta_2}{\partial x_1\partial x_2}\Big)=0,\quad \text{at} \  y=0;\label{equ: boundaryII-5}\\
&\frac{\partial \phi_5}{\partial y}=0,\quad \text{at} \ y=-1.\label{equ: boundaryIII-5}
\end{flalign}
Introducing the expressions previously obtained for $\phi_1,\phi_2,\phi_3,\phi_4$ into \eqref{equ: poisson5}, we have
\begin{equation*}
\begin{split}
  \frac{\partial^2 \phi_5}{\partial x_0^2}+\frac{\partial^2 \phi_5}{\partial y^2}&=\frac{2iw(y+1)}{\sinh k}\sinh [k(y+1)]\frac{\partial^2 A}{\partial x_1^2}e^{i\theta}-\frac{2w}{\sinh k}\cosh [k(y+1)]\frac{\partial B}{\partial x_1}e^{i\theta}\\
  &\quad-\frac{2w}{\sinh k}\cosh [k(y+1)]\frac{\partial A}{\partial x_2}e^{i\theta}
+\frac{iw}{k\sinh k}\cosh [k(y+1)]\frac{\partial^2 A}{\partial x_1^2}e^{i\theta}+c.c.\\
&\quad\quad+\frac{1}{2}\frac{\partial^4 \psi_1}{\partial x_1^4}(y+1)^2-\frac{\partial^2 \psi_3}{\partial x_1^2}-2\frac{\partial^2 \psi_2}{\partial x_1\partial x_2}-\frac{\partial^2 \psi_1}{\partial x_2^2}-2\frac{\partial^2 \psi_1}{\partial x_1\partial x_3}.\\
  \end{split}
\end{equation*}
This has a solution given by the following form
\begin{equation*}\label{solu:phi5}
\begin{split}
  \phi_5&=-\frac{w(y+1)}{k\sinh k}\sinh [k(y+1)]\frac{\partial A}{\partial x_2}e^{i\theta}+\frac{iw(y+1)^2}{2k\sinh k}\cosh [k(y+1)]\frac{\partial^2 A}{\partial x_1^2}e^{i\theta}\\
  &-\frac{w(y+1)}{k\sinh k}\sinh [k(y+1)]\frac{\partial B}{\partial x_1}e^{i\theta}+\frac{-iw}{k\sinh k}\cosh [k(y+1)]\widetilde{B}e^{i\theta}+c.c.\\
  &\frac{1}{24}\frac{\partial^4 \psi_1}{\partial x_1^4}(y+1)^4-\frac{1}{2}\frac{\partial^2 \psi_3}{\partial x_1^2}(y+1)^2-\frac{\partial^2 \psi_2}{\partial x_1\partial x_2}(y+1)^2-\frac{1}{2}\frac{\partial^2 \psi_1}{\partial x_2^2}(y+1)^2-\frac{\partial^2 \psi_1}{\partial x_1\partial x_3}(y+1)^2+\psi_5\\
  \end{split}
\end{equation*}
where $\widetilde{B}\in \mathbb{C}$ (the homogeneous solution) and $\psi_5\in\mathbb{R}$ are arbitrary functions independent of the slow variables $x_0,t_0,y$.
Then, it follows from \eqref{equ: boundaryI-5} that
\begin{equation*}
\begin{split}
  \frac{\partial \eta_5}{\partial t_0}&= \frac{\partial \phi_5}{\partial y}-\frac{\partial \eta_4}{\partial t_1}-\frac{\partial \eta_3}{\partial t_2}-\frac{\partial \eta_2}{\partial t_3}-\frac{\partial \phi_1}{\partial x_1}\frac{\partial \eta_2}{\partial x_1}\\
  &=\Big\{-\frac{w}{k}(1+k\coth k)\frac{\partial A}{\partial x_2}e^{i\theta}+\frac{iw}{2k}(k+2\coth k)\frac{\partial^2 A}{\partial x_1^2}e^{i\theta}-\frac{w}{k}(1+k\coth k)\frac{\partial B}{\partial x_1}e^{i\theta}\\
  &\quad-iw\widetilde{B}e^{i\theta}
  +\frac{1}{6}\frac{\partial^4 \psi_1}{\partial x_1^4}-\frac{\partial^2 \psi_3}{\partial x_1^2}-2\frac{\partial^2 \psi_2}{\partial x_1\partial x_2}-\frac{\partial^2 \psi_1}{\partial x_2^2}-2\frac{\partial^2 \psi_1}{\partial x_1\partial x_3}\Big\}\\
  &-\Big\{-\frac{i}{k}(1+k\coth k)\frac{\partial^2 A}{\partial x_1\partial t_1}e^{i\theta}+\frac{\partial B}{\partial t_1}e^{i\theta}-\frac{i}{w}\frac{\partial^2 A}{\partial t_1^2}e^{i\theta}+\frac{\partial \xi_4}{\partial t_1}\Big\}\\
  &-\Big\{\frac{\partial A}{\partial t_2}e^{i\theta}+\frac{\partial \xi_3}{\partial t_2}\Big\}-\frac{\partial \xi_2}{\partial t_3}-\frac{\partial \psi_1}{\partial x_1}\frac{\partial \xi_2}{\partial x_1}+ c.c..\\
  \end{split}
\end{equation*}
Integrating the above equation with respect to the slow variable $t_0$, we obtain
\begin{equation*}
\begin{split}
\eta_5
  &=\Big\{-\frac{i}{k}(1+k\coth k)\frac{\partial A}{\partial x_2}e^{i\theta}-\frac{1}{2k}(k+2\coth k)\frac{\partial^2 A}{\partial x_1^2}e^{i\theta} -\frac{i}{k}(1+k\coth k)\frac{\partial B}{\partial x_1}e^{i\theta}+\widetilde{B}e^{i\theta}
  \Big\}\\
  &-\Big\{\frac{1}{wk}(1+k\coth k)\frac{\partial^2 A}{\partial x_1\partial t_1}e^{i\theta}+\frac{i}{w}\frac{\partial B}{\partial t_1}e^{i\theta}+\frac{1}{w^2}\frac{\partial^2 A}{\partial t_1^2}e^{i\theta}\Big\} -\frac{i}{w}\frac{\partial A}{\partial t_2}e^{i\theta}+\xi_5 +c.c.,\\
\end{split}
\end{equation*}
provided that the non-secularity condition
\begin{equation}\label{relation10}
\frac{1}{6}\frac{\partial^4 \psi_1}{\partial x_1^4}-\frac{\partial^2 \psi_3}{\partial x_1^2}-2\frac{\partial^2 \psi_2}{\partial x_1\partial x_2}-\frac{\partial^2 \psi_1}{\partial x_2^2}-2\frac{\partial^2 \psi_1}{\partial x_1\partial x_3}-\frac{\partial \xi_4}{\partial t_1}-
\frac{\partial \xi_3}{\partial t_2}-\frac{\partial \xi_2}{\partial t_3}-\frac{\partial \psi_1}{\partial x_1}\frac{\partial \xi_2}{\partial x_1}=0
\end{equation}
is satisfied.
Eliminating of $\xi_2,\xi_3,\xi_4$ from \eqref{relation10}, we obtain the following equation
\begin{equation}\label{relation13}
\begin{split}
  (\frac{1}{W}-\frac{1}{3})\frac{\partial^4 \psi_1}{\partial x_1^4}&+(\frac{\partial^4 \psi_3}{\partial t_1^2}-\frac{\partial^4 \psi_3}{\partial x_1^2})+2(\frac{\partial^2 \psi_2}{\partial t_1\partial t_2}-\frac{\partial^2 \psi_2}{\partial x_1\partial x_2})+2(\frac{\partial^2 \psi_1}{\partial t_1\partial t_3}-\frac{\partial^2 \psi_1}{\partial x_1\partial x_3})\\
  &+(\frac{\partial^2 \psi_1}{\partial t_2^2}-\frac{\partial^2 \psi_1}{\partial x_2^2})+2\frac{\partial \psi_1}{\partial x_1}\frac{\partial^2\psi_1}{\partial x_1\partial t_1}=0.
\end{split}
\end{equation}
Now, we make the assumption that the functions $\psi_i$ are given by the special forms $\psi_i(x_1-t_1, x_2-t_2,x_3,t_3,\cdots)$ for $i=1,2,3,$ that is, we consider the long waves that propagate only in the positive direction.  Upon letting $u=\frac{\partial \psi_1}{\partial x_1}$, the assumption above implies that
$u=-\frac{\partial \psi_1}{\partial t_1}$ and thus the equation \eqref{relation13} reduces to
\begin{equation}\label{KdV}
\frac{\partial u}{\partial t_3}+\frac{\partial u}{\partial x_3}
  +u\frac{\partial u}{\partial x_1}+(\frac{1}{6}-\frac{1}{2W})\frac{\partial^3 u}{\partial x_1^3}=0,
\end{equation}
which is the \textit{pure} KdV equation with no coupling terms.
Inserting $\phi_5$ and $\eta_5$ into \eqref{equ: boundaryII-5}, we have
\begin{equation*}
\begin{split}
  &\Big\{\frac{iw^2}{k}\frac{\partial A}{\partial x_2}e^{i\theta}+\frac{w^2}{2k\tanh k}\frac{\partial^2 A}{\partial x_1^2}e^{i\theta}
  +\frac{iw^2}{k}\frac{\partial B}{\partial x_1}e^{i\theta}+\frac{-w^2}{k\tanh k}\widetilde{B}e^{i\theta}\Big\}\\
  &+\Big\{-\frac{w}{k}\frac{\partial^2 A}{\partial x_1\partial t_1}e^{i\theta}+\frac{-iw}{k\tanh k}\frac{\partial B}{\partial t_1}e^{i\theta}-\frac{1}{2}\frac{\partial^3\psi_2}{\partial x_1^2\partial t_1}-\frac{\partial^3 \psi_1}{\partial x_1\partial x_2\partial t_1}+\frac{\partial \psi_4}{\partial t_1}\Big\}\\
  &+\frac{-iw}{k\tanh k}\frac{\partial A}{\partial t_2}e^{i\theta}-\frac{1}{2}\frac{\partial^3 \psi_1}{\partial x_1^2\partial t_2}+\frac{\partial \psi_3}{\partial t_2}+\frac{\partial \psi_2}{\partial t_3}+\frac{\partial \psi_1}{\partial t_4}+\frac{\partial \psi_1}{\partial x_1}\Big[\frac{\partial \psi_1}{\partial x_2}+\frac{\partial \psi_2}{\partial x_1}+\frac{w}{\tanh k}Ae^{i\theta}\Big]\\
  &+(1+\frac{k^2}{W})\Big\{-\frac{i}{k}(1+k\coth k)\frac{\partial A}{\partial x_2}e^{i\theta}-\frac{1}{2k}(k+2\coth k)\frac{\partial^2 A}{\partial x_1^2}e^{i\theta}-\frac{i}{k}(1+k\coth k)\frac{\partial B}{\partial x_1}e^{i\theta}+\widetilde{B}e^{i\theta}
  \\
  &-\Big[\frac{1}{wk}(1+k\coth k)\frac{\partial^2 A}{\partial x_1\partial t_1}e^{i\theta}+\frac{i}{w}\frac{\partial B}{\partial t_1}e^{i\theta}+\frac{1}{w^2}\frac{\partial^2 A}{\partial t_1^2}e^{i\theta}\Big]
  -\frac{i}{w}\frac{\partial A}{\partial t_2}e^{i\theta}\Big\}\\&-\frac{1}{W}\Big\{2[(1+k\coth k)\frac{\partial^2 A}{\partial x_1^2}e^{i\theta}+ik\frac{\partial B}{\partial x_1}e^{i\theta}
+\frac{k}{w}\frac{\partial^2 A}{\partial t_1\partial x_1}e^{i\theta}]\\
&+2ik\frac{\partial A}{\partial x_2}e^{i\theta}+\frac{\partial^2 A}{\partial x_1^2}e^{i\theta}+\frac{\partial^2 \xi_3}{\partial x_1^2}+2\frac{\partial^2 \xi_2}{\partial x_1\partial x_2}\Big\}+\xi_5 + c.c.=0.
  \end{split}
\end{equation*}
After eliminating of the term $\widetilde{B}$ due to the dispersion relation \eqref{relation}, we obtain
\begin{equation*}\label{relation14}
\begin{split}
  &\Big\{\frac{iw^2}{k}\frac{\partial A}{\partial x_2}e^{i\theta}+\frac{w^2}{2k\tanh k}\frac{\partial^2 A}{\partial x_1^2}e^{i\theta}
  +\frac{iw^2}{k}\frac{\partial B}{\partial x_1}e^{i\theta}\Big\}+\frac{-iw}{k\tanh k}\frac{\partial A}{\partial t_2}e^{i\theta}-\frac{1}{2}\frac{\partial^3 \psi_1}{\partial x_1^2\partial t_2}+\frac{\partial \psi_3}{\partial t_2}+\frac{\partial \psi_2}{\partial t_3}+\frac{\partial \psi_1}{\partial t_4}\\
  &+\Big\{-\frac{w}{k}\frac{\partial^2 A}{\partial x_1\partial t_1}e^{i\theta}+\frac{-iw}{k\tanh k}\frac{\partial B}{\partial t_1}e^{i\theta}-\frac{1}{2}\frac{\partial^3\psi_2}{\partial x_1^2\partial t_1}-\frac{\partial^3 \psi_1}{\partial x_1\partial x_2\partial t_1}+\frac{\partial \psi_4}{\partial t_1}\Big\}\\
  &+(1+\frac{k^2}{W})\Big\{-\frac{i}{k}(1+k\coth k)\frac{\partial A}{\partial x_2}e^{i\theta}-\frac{1}{2k}(k+2\coth k)\frac{\partial^2 A}{\partial x_1^2}e^{i\theta}-\frac{i}{k}(1+k\coth k)\frac{\partial B}{\partial x_1}e^{i\theta}
  \\
  &-\Big[\frac{1}{wk}(1+k\coth k)\frac{\partial^2 A}{\partial x_1\partial t_1}e^{i\theta}+\frac{i}{w}\frac{\partial B}{\partial t_1}e^{i\theta}+\frac{1}{w^2}\frac{\partial^2 A}{\partial t_1^2}e^{i\theta}\Big]
  -\frac{i}{w}\frac{\partial A}{\partial t_2}e^{i\theta}\Big\}\\&-\frac{1}{W}\Big\{2[(1+k\coth k)\frac{\partial^2 A}{\partial x_1^2}e^{i\theta}+ik\frac{\partial B}{\partial x_1}e^{i\theta}
  +\frac{k}{w}\frac{\partial^2 A}{\partial t_1\partial x_1}e^{i\theta}] +\frac{\partial \psi_1}{\partial x_1}\Big[\frac{\partial \psi_1}{\partial x_2}+\frac{\partial \psi_2}{\partial x_1}+\frac{w}{\tanh k}Ae^{i\theta}\Big]\\
  &+2ik\frac{\partial A}{\partial x_2}e^{i\theta}+\frac{\partial^2 A}{\partial x_1^2}e^{i\theta}+\frac{\partial^2 \xi_3}{\partial x_1^2}+2\frac{\partial^2 \xi_2}{\partial x_1\partial x_2}\Big\}+\xi_5 +c.c.=0.\\
  \end{split}
\end{equation*}
Moreover, the terms involved $\frac{\partial A}{\partial x_2},\frac{\partial A}{\partial t_2}$ and
$\frac{\partial B}{\partial x_1},\frac{\partial B}{\partial t_1}$ can be rewritten in compact forms similar to the equation \eqref{non-seculartiy2}, we therefore have
\begin{equation}\label{relation16}
\begin{split}
  &\frac{w^2}{2k\tanh k}\frac{\partial^2 A}{\partial x_1^2}e^{i\theta}
  -\frac{w}{k}\frac{\partial^2 A}{\partial x_1\partial t_1}e^{i\theta}-\frac{1}{2}\frac{\partial^3\psi_2}{\partial x_1^2\partial t_1}-\frac{\partial^3 \psi_1}{\partial x_1\partial x_2\partial t_1}+\frac{\partial \psi_4}{\partial t_1}\\
  &-\frac{1}{2}\frac{\partial^3 \psi_1}{\partial x_1^2\partial t_2}+\frac{\partial \psi_3}{\partial t_2}+\frac{\partial \psi_2}{\partial t_3}+\frac{\partial \psi_1}{\partial t_4}
  +\frac{\partial \psi_1}{\partial x_1}\Big[\frac{\partial \psi_1}{\partial x_2}+\frac{\partial \psi_2}{\partial x_1}+\frac{w}{\tanh k}Ae^{i\theta}\Big]\\
  &+(1+\frac{k^2}{W})\Big\{-\frac{1}{2k}(k+2\coth k)\frac{\partial^2 A}{\partial x_1^2}e^{i\theta}-\Big[\frac{1}{wk}(1+k\coth k)\frac{\partial^2 A}{\partial x_1\partial t_1}e^{i\theta}+\frac{1}{w^2}\frac{\partial^2 A}{\partial t_1^2}e^{i\theta}\Big]
  \Big\}\\
  &-\frac{1}{W}\Big\{2[(1+k\coth k)\frac{\partial^2 A}{\partial x_1^2}e^{i\theta}
  +\frac{k}{w}\frac{\partial^2 A}{\partial t_1\partial x_1}e^{i\theta}]
  +\frac{\partial^2 A}{\partial x_1^2}e^{i\theta}+\frac{\partial^2 \xi_3}{\partial x_1^2}+2\frac{\partial^2 \xi_2}{\partial x_1\partial x_2}\Big\}\\& -\frac{2iw}{k\tanh k}\Big[\frac{\partial A}{\partial t_2}+V_g\frac{\partial A}{\partial x_2}\Big]-\frac{2iw}{k\tanh k}\Big[\frac{\partial B}{\partial t_1}+V_g\frac{\partial B}{\partial x_1}\Big]+\xi_5 +c.c.=0.\\
  \end{split}
\end{equation}
Now, from the equation \eqref{non-seculartiy2} we deduce that
\begin{equation}
\frac{\partial^2 A}{\partial x_1\partial t_1}=-V_g\frac{\partial^2 A}{\partial x_1^2},
\quad
\frac{\partial^2 A}{\partial t_1^2}=V_g^2\frac{\partial^2 A}{\partial x_1^2}.
\end{equation}
Plugging them into the equation \eqref{relation16},
we then have the following equations
\begin{equation}\label{relation17}
\begin{split}
  -\frac{2iw}{k\tanh k}\Big[\frac{\partial B}{\partial t_1}+V_g\frac{\partial B}{\partial x_1}\Big]
  -\frac{2iw}{k\tanh k}\Big[\frac{\partial A}{\partial t_2}+V_g\frac{\partial A}{\partial x_2}\Big]
  +C(k)\frac{\partial^2 A}{\partial x_1^2}+\frac{w}{\tanh k}A\frac{\partial \psi_1}{\partial x_1}=0,
  \end{split}
\end{equation}
where
\begin{equation*}
  \begin{split}
    C(k)&=\frac{-1}{k\tanh k}V_g^2+V_g\Big\{\frac{w}{k}+\frac{w}{k^2\tanh k}(1+k\coth k)+\frac{2k}{wW}\Big\}\\&-\frac{w^2}{k^2\tanh^2 k}-\frac{3+2k\coth k}{W};
  \end{split}
\end{equation*}
and
\begin{equation*}
\begin{split}
&-\frac{1}{2}\frac{\partial^3\psi_2}{\partial x_1^2\partial t_1}-\frac{\partial^3 \psi_1}{\partial x_1\partial x_2\partial t_1}+\frac{\partial \psi_4}{\partial t_1}-\frac{1}{2}\frac{\partial^3 \psi_1}{\partial x_1^2\partial t_2}+\frac{\partial \psi_3}{\partial t_2}+\frac{\partial \psi_2}{\partial t_3}+\frac{\partial \psi_1}{\partial t_4}\\
&-\frac{1}{W}[\frac{\partial^2 \xi_3}{\partial x_1^2}+2\frac{\partial^2 \xi_2}{\partial x_1\partial x_2}]+\xi_5+\frac{\partial \psi_1}{\partial x_1}\Big[\frac{\partial \psi_1}{\partial x_2}+\frac{\partial \psi_2}{\partial x_1}\Big]=0.
\end{split}
\end{equation*}
The equation \eqref{relation17} is the linear Schr\"odinger equation. \\

\noindent \textbf{Remark 4.}  It is therefore established that both the \textit{linear} Schr\"odinger \eqref{relation17} and KdV \eqref{KdV} equations appear at the same order $O(\epsilon^5)$.  However, there is \textit{no} coupling terms in the KdV equation.

\vskip .1in
\noindent{\bf Acknowledgement.}  The authors would like to thank Bernard Deconinck and Benjamin L. Segal (the University of Washington) and Shu-ming Sun (Virginia Tech) for valuable conversations. The first author is supported by NSFC-11101171 and gratefully acknowledges financial support from China Scholarship Council. Also, he would like to thank the Department of Mathematics and Statistics, USU, for hosting his visit during which this work was carried out.
\bibliographystyle{siam}

\begin{thebibliography}{1}



\bibitem{albert}
{\sc J. Albert and S. Bhattarai}, \textit{Existence and stability of a two-parameter family of solitary waves for an NLS-KdV system}, Adv. Diff. Eqns., Vol. 18 (2013), 1129-1164.

\bibitem{pava1}
{\sc J. Angulo Pava}, {\em Stability of solitary wave solutions for equations of short and long dispersive waves}, Elec. Jour. Diff. Eqns., Vol. 72 (2006), 1-18.

\bibitem{pava2}
{\sc J. Angulo Pava, C. Matheus and D. Pilod}, {\em Global well-posedness and nonlinear stability of periodic traveling waves for a Schr\"odinger-Benjamin-Ono system}, Comm. Pure and Applied Anal., Vol. 8 (2009), 3, 815-844.

\bibitem{appert}
{\sc K. Appert and J. Vaclavik}, {\em Dynamics of coupled solitons}, Phys. Fluids, Vol. 20 (1977), 1845-1849.



\bibitem{chen}
{\sc L. Chen}, {\em Orbital stability of solitary waves of the nonlinear Schr\"odinger-KdV equation}, Jour. Partial Diff. Eqns., Vol 12 (1999), 11-25.

\bibitem{corcho}
{\sc A.J. Corcho and F. Linares}, {\em Well-posedness for the Schr\"odinger-Korteweg-de Vries system}, Trans. Amer. Math. Soc., Vol. 359 (2007), 4089-4106.

\bibitem{DNS}
{\sc B. Deconinck, N.V. Nguyen and B.L. Segal}, {\em The interaction of long and short waves in dispersive media}, Pre-print.



\bibitem{dias}
{\sc J.P. Dias, M. Figueira and F. Oliveira}, {\em Well-posedness and existence of bound states for a coupled Schr\"odinger-gKdV system}, Nonlinear Anal., Vol. 73 (2010), 2686-2698.

\bibitem{funakoshi}
{\sc M. Funakoshi and M. Oikawa}, {\em The resonant interactions between a long internal gravity wave and a surface gravity wave packet}, Jour. Phys. Soc. Japan, Vol. 52 (1983), 1982-1995.

\bibitem{ggkm}
{\sc C. S. Gardner, J. M. Greene, M. D. Kruskal and R. M. Miura}, {\em Method for solving the Korteweg-deVries equation}, Phys. Rev. Lett., Vol. 19 (1967), 1095-1097.

\bibitem{ikezi}
{\sc H. Ikezi, K. Nishikawa, H. Hojo and K. Mima}, {\em Coupled electron-plasma and ion-acoustic solitons excited by parametric instability}, Proc. 5th Intern. Conf. on Plasma Phys. and Contr. Nucl. Fusion Res., Tokyo (1974), 239-248.



\bibitem{KSK}
{\sc T. Kawahara, N. Sugimoto and T. Kakutani}, {\em Nonlinear interaction between short and long capillary-gravity waves}, Jour. Phys. Soc. Japan, Vol. 39, 5 (1975), 1379-1386.

\bibitem{nishikawa}
{\sc K. Nishikawa, H. Hojo, K. Mima and H. Ikezi}, {\em Coupled nonlinear electron-plasma and ion-acoustic waves}, Phys. Rev. Lett., Vol. 33 (1974), 148-151.



\bibitem{sz}
{\sc A. shabat and V. Zakharov}, {\em Exact theory of two-dimensional self-focusing and one-dimensional self-modulation of waves in nonlinear media}, Sov. Phys JETP, Vol. 34 (1972), 62-69.



\bibitem{Zak}
{\sc V. E. Zakharov}, {\em Collapse of Langmuir waves}, Sov. Phys. JETP., Vol. 35 (1972), 908-914.

\end{thebibliography}

\end{document}